\documentclass[12pt,english,dvips]{scrartcl}
\usepackage{amsmath}
\usepackage[T1]{fontenc}
\usepackage[ansinew]{inputenc}

\begin{document}

\begin{titlepage} \vspace{0.2in} 

\begin{center} {\LARGE \bf 

On the Kaluza-Klein geometrization\\ of the Electro-Weak Model\\ within a gauge theory of the\\ 5-dimensional Lorentz group}\\ 
\vspace*{0.8cm}
{\bf Orchidea Maria Lecian}\\ \vspace*{0.3cm}
{\bf Giovanni Montani}\\ \vspace*{1cm}
ICRA---International Center for Relativistic Astrophysics\\
Dipartimento di Fisica (G9),\\
Universit\`a  di Roma, ''La Sapienza'',\\
Piazzale Aldo Moro 5, 00185 Rome, Italy.\\
e-mail: lecian@icra.it, montani@icra.it\\
\vspace*{1.8cm}
 \vspace*{1cm} \\

{\bf   Abstract  \\ } \end{center} \indent
The geometrization of the Electro-Weak Model is achieved in a 5-dimensional Riemann-Cartan framework. Matter spinorial fields are extended to 5 dimensions by the choice of a proper dependence on the extra-coordinate and of a normalization factor. U(1) weak hyper-charge  gauge fields are obtained from a Kaluza-Klein scheme, while  the  tetradic projections of the extra-dimensional contortion fields are interpreted as SU(2) weak isospin gauge fields. SU(2) generators are derived by the identification of the weak isospin current to the extra-dimensional current term in the Lagrangian density of the local Lorentz group. The geometrized U(1) and SU(2) groups will provide the proper transformation laws for bosonic and spinorial fields. Spin connections will be found to be purely Riemannian.
\end{titlepage}

\section*{Introduction}

The search after unification has always been a constant effort in the history of modern Physics, and gauge theories are the tool which has favored its development: in the Standard Model, all the features of matter fields can be ascribed to the gauge symmetries that define the model itself. Gauge theories describe successfully all interactions but gravity, whose phenomenology is explained by the characteristics of the space-time: among several attempts to unify such different aspects in a coherent system, Lorentz gauge theories allow to express the principle of equivalence as an internal symmetry of the space-time.\\
In this paper we'll try to geometrize the Standard Electro-Weak Model, and further degrees of freedom will be looked for in an extra-dimensional background. The Electro-Weak Model consists of two symmetries groups, the U(1) weak hyper-charge Abelian group and the SU(2) weak isospin non-Abelian group. The former will be geometrized in a 5-dimensional Kaluza-Klein frame-work, where the gauge field is identified in the non-diagonal components of the 5-dimensional metric tensor, while the latter will be searched for in the tetradic projection of the pertinent components of the contortion field. Spinorial fields can be described in such a scenario by their extension to 5 dimensions.
\\ Most of the properties of the objects defined in this paper will be suggested by the basic features of Kaluza-Klein theories: in such a scenario, five dimensions are considered. Gauge interactions can be geometrized by means of some coordinate transformations, i.e. extra-dimensional translations only are allowed. The structure of the space-time is, for the ground state, not a generic 5-dimensional manifold, but the direct sum of a generic 4-dimensional one and a ring. The extension of general relativity to 5 dimensions within this scheme leads to the violation of the principle of general relativity and of the principle of equivalence. Nonetheless, ''{\it the existence of the ground state which does not have the form of $M^{5}$ can be best understood as an effect of spontaneous symmetry breaking}'', \cite{bla}. \\
In the first section, a review of Kaluza-Klein theories will be presented \cite{K21},\cite{K26}, \cite{K26a},\cite{ACF87}, \cite{wit}: it will be shown how it properly fits weak hyper-charge U(1) group, and chiral states will arise naturally from a 5-dimensional Dirac equation \cite{bla}, \cite{bai}, \cite{W83}. Gauge fields will be introduced in the metric tensor, and the tetradic projection of the derivative will provide the right coupling of matter fields and the bosonic field \cite{bla}. The dependence on the fifth coordinate \cite{mar} will provide U(1) weak hyper-charge transformations for spinorial fields \cite{fra}, \cite{bjo}. \\
The second section is aimed to develop a 4-dimensional gauge theory of the Lorentz group \cite{mermon}:
starting from the analogy with a generic gauge theory, the theory will be worked out in a Riemann-Cartan space \cite{goe}, \cite{ten}, and variation with respect to the objects involved in the total action will lead to standard equations \cite{wei}.  \\
These results will be extended  to 5 dimensions in the third section \cite{pav1}, \cite{pav2}, : such an extension will require care, because of the structure of the manifold $V^{5}=V^{4}\oplus S^{1}$ considered in Kaluza-Klein theories. The 5-dimensional Lorentz group, and its gauge fields $A^{\bar{A}\bar{B}}_{\Omega}$, will split up naturally in the ordinary 4-dimensional Lorentz group and SU(2) weak isospin non-Abelian group \cite{mand}, \cite{jo}, whose generators will be defined by the recognition of the pertinent Lagrangian term to the weak isospin current of the Standard Model. Structure constants  will provide the proper gauge transformation laws  for 4-dimensional Lorentz fields and for SU(2) fields \cite{MS84}, \cite{bjo}. World transformation laws will request that $A^{\bar{A}\bar{B}}_{5}$ fields vanish; these fields, however, had no precise role in this model. Lorentz, U(1) and SU(2) gauge transformations will also be established for spinorial fields, and, a formal difference with the Standard Model notwithstanding, they will be explained to be coherent to standard  U(1)$\otimes$SU(2).\\
In the fourth section, the previous results will be collected together. The tetradic projection of spin and gauge connections will be evaluated \cite{gio}, \cite{fra}, \cite{mar}, \cite{bla}, \cite{vac}, \cite{det}: spin connections will be found to be purely Riemannian, and gauge connections will come directly from the preceding calculations. By means of dimensional reduction, the Standard Electro-Weak Model is therefore restored \cite{MS84}, \cite{kum}.\\
Brief concluding remarks follow.\\
In this paper we have chosen the following conventions:\\
- small Greek letters for 4-dimensional world indices, $\mu=0,1,2,3$;\\
- capital Greek letters for 5-dimensional world indices, $\Omega=0,1,2,3,5$;\\
- small Latin letters for 4-dimensional tetradic indices, $\bar{a}=0,1,2,3$;\\
- capital Latin letters for 5-dimensional tetradic indices, $\bar{A}=0,1,2,3,5$;\\
- $i,j,k$ for $1,2,3$;\\ 
- natural units, $\hbar \equiv c \equiv 1$.

\section{Kaluza-Klein theories: local hyper-charge U(1) transformations and chirality}

In a 5-dimensional Kaluza-Klein scheme, it is possible to geometrize an Abelian gauge group \cite{K21}, \cite{K26}, \cite{K26a},\cite{ACF87}; as far as the Standard Electro-Weak Model is concerned, it sounds sensible to identify such a gauge field to the weak hyper-charge U(1) one, $B_{\mu}$.
Let 
\begin{equation}\label{struttura}
V^{5}=V^{4}\oplus S^{1}
\end{equation} 
be a 5-dimensional $C^{\infty}$ manifold, direct sum of a generic 4-dimensional manifold and a ring. Because of this structure, the metric tensor $j_{AB}$ has to be a periodic function of the fifth coordinate, $x^{5}$, so that
\begin{equation}
j_{AB}(x^{\mu},x^{5})=j_{AB}(x^{\mu},x^{5}+L),
\end{equation}
$L$ being the length of the ring, and can therefore be Fourier expanded. For the cylindrical hypothesis, which states that every function with physical relevance must not depend on the fifth coordinate, the expansion has to be truncated at $0$-order.\\
Allowed coordinate transformations are obtained by the constraint that $j_{55}$ be a scalar ( for our purposes we can assume since now $j_{55}=1$, as we are not interested in the dynamics of this scalar field  ) and by the cylindrical hypothesis, which implies that the transformation law for the metric tensor does not depend on $x^{5}$: we finally get
\begin{equation}\label{ammesse}
x'^{\mu}=x'^{\mu}(x^{\rho});x'^{5}=x^{5}+\alpha (x^{\rho}).
\end{equation} 
Therefore, the components of the metric tensor are
\begin{equation}\label{basse}
\begin{cases}
j_{55}=1\cr
j_{5\mu}=g'k\tilde{B}_{\mu}\cr
j_{\mu\nu}-(\tilde{g}'k)^{2}\tilde{B}_{\mu}\tilde{B}_{nu}=g_{\mu\nu},\cr
\end{cases}
\end{equation}
where $k$ is a constant, which we need to introduce for dimensional reasons, $\tilde{g}'$ and $\tilde{B}_{\mu}$ can be interpreted as a coupling constant and a gauge field, respectively, which are related to the weak hyper-charge constant and the weak hyper-charge gauge field, as it'll be worked out later on. After standard manipulation one gets the inverse components of the metric,
\begin{equation}
\begin{cases}
j^{55}=1+(\tilde{g}'k)^{2}\tilde{B}_{\sigma}\tilde{B}^{\sigma}\cr
j^{5\mu}=-\tilde{g}'k\tilde{B}^{\mu}\cr
j^{\mu\nu}=g^{\mu\nu}.\cr
\end{cases}
\end{equation}
From (\ref{ammesse}) and(\ref{basse}) it is clear that $B_{\mu}$ transforms like an ordinary 4-vector if we consider pure 4-dimensional transformations,
\begin{equation}
\tilde{B}_{\mu}\rightarrow \tilde{B}_{\nu}\frac{\partial x^{\nu}}{\partial x'^{\mu}},
\end{equation}
while it behaves like a gauge field for the second of (\ref{ammesse}), provided that $\alpha(x^{\rho})=\tilde{g}'k\alpha'(x^{\rho})$, so that
\begin{equation}\label{alfa}
\tilde{B}_{\mu}\rightarrow \tilde{B}_{\mu}+\partial_{\mu}\alpha'(x^{\sigma}).
\end{equation}
Tetradic vectors are found by imposing
\begin{equation}
j_{AB}=V^{\bar{A}}_{A}V^{\bar{B}}_{B}\eta_{\bar{A}\bar{B}};
\end{equation}
with straightforward calculation one finds 
\begin{equation}
\begin{cases}
V_{5}^{\bar{\mu}}=0\cr V_{5}^{\bar{5}}=1\cr V^{\bar{5}}_{\mu}=\tilde{g}'k \tilde{B}_{\mu}\cr V_{\mu}^{\bar{\mu}} : g_{\mu\nu}=V_{\mu}^{\bar{\mu}}V_{\nu}^{\bar{\nu}}\eta_{\bar{\mu}\bar{\nu}},\cr
\end{cases}
\end{equation}
and
\begin{equation}
\begin{cases}
V^{\mu}_{\bar{5}}=0\cr
V^{5}_{\bar{5}}=1\cr
V^{5}_{\bar{\mu}}=-\tilde{g}'k\tilde{B}_{\mu}V^{\mu}_{\bar{\mu}}\cr
V^{\mu}_{\bar{\mu}}  : g^{\mu\nu}=\eta^{\bar{\mu}\bar{\nu}}V^{\mu}_{\bar{\mu}}V^{\nu}_{\bar{\nu}}.\cr
\end{cases}
\end{equation}
It can be shown that the action $^{5}S$ splits up into the ordinary Einstein action and the action of an Abelian gauge field 
\begin{equation}\label{azionefinalmentedue}
S=\frac{-c^{3}}{16 \pi  {^{5}G}}\int dx^{5}d^{4}x\sqrt{-j}{^{5}R}=
\frac{-c^{3}}{16 \pi  {^{5}G}}\int d^{4}x\sqrt{-g}[^{4}R+(\frac{\tilde{g}'k}{2})^{2}\tilde{F}_{\mu\nu}\tilde{F}^{\mu\nu}],
\end{equation}
where dimensional reduction has been performed,   
\begin{equation}
\tilde{F}_{\mu\nu} \equiv \partial_{\nu}\tilde{B}_{\mu}-\partial_{\mu}\tilde{B}_{\nu},
\end{equation}
is the bosonic Lagrangian density of the field $\tilde{B}_{\mu}$, and ${^{5}G}$ is related to the 4-dimensional Newton constant by $G=\frac{{^{5}G}}{L}$.\\
Coupling with matter is an important aspect of this article. In the original work by Kaluza it had not been taken into account, and many efforts have been made in order to find a proper description.\\
For a massless fermionic field, the Lagrangian density reads
\begin{equation}\label{13}
L=-\frac{i}{2}\bar{\psi_{i}}\gamma^{\mu}\partial_{\mu}\psi_{i} + H.C.,
\end{equation}
where $H.C.$ denotes Hermitian conjugation; its immediate 5-dimensional extensions could be
\begin{equation}\label{festa}
L=-\frac{i}{2}\bar{\chi_{i}}\gamma^{A}\partial_{A}\chi_{i} +H.C.=-\frac{i}{2}\bar{\chi_{i}}\gamma^{\bar{A}}\partial_{\bar{A}}\chi_{i}+H.C.,
\end{equation}
where  $\chi_{i}$ is a generic 5-dimensional fermionic field and $\gamma^{5}$ the fifth Dirac matrix, defined as $\gamma^{5}=i\gamma^{0}\gamma^{1}\gamma^{2}\gamma^{3}$. $\gamma^{\bar{5}}$ is still a good Dirac matrix, as it anticommutes whit all the other $\gamma^{\bar{\mu}}$'s in 4 dimensions, and, in 5 dimensions, we have
\begin{equation}\label{gamma}
[\gamma^{\bar{A}},\gamma^{\bar{B}}]_{+}=2I\eta^{\bar{A}\bar{B}}.
\end{equation}
A Lagrangian density has to be invariant under Lorentz transformations: the 5-  dimensional Lagrangian density (\ref{festa}) will be shown to be invariant under 5-dimensional local Lorentz transformations as they will be defined, in a Kaluza-Klein framework, in the next sections.\\ 
The structure of the manifold we are considering suggests a factorization of the wave function as a consequence of the direct sum in (\ref{struttura}), \cite{mar}, so that we can guess
\begin{equation}
{^{5}\chi}_{i}\equiv {^{5}\chi}_{i}(x^{\rho},x^{5})\propto {^{4}\psi}_{i}(x^{\rho})f(x^{5});
\end{equation}
moreover, since the fifth dimension is a ring, the functional dependence on $x^{5}$ can be chosen as a complex phase,
\begin{equation}
f(x^{5})=e^{\frac{i2\pi N_{i}x^{5}}{L}},
\end{equation}
whose periodicity is this way connected to the length $L$ of the ring by the integer dimensionless parameter $N_{i}$, which is related to the weak hyper-charge eigenvalue, and describes different ''particles'' according to different properties of their ''motion'' along the extra-dimension. We can append a suitable renormalization factor, whose validity will be tested afterwords, and the 5-spinor finally reads 
\begin{equation}\label{chi}
{^{5}\chi}_{i}(x^{\rho},x^{5})=\frac{1}{\sqrt{L}}{^{4}\psi}_{i}(x^{\rho})e^{\frac{i2\pi N_{i}x^{5}}{L}}.
\end{equation}
In a Kaluza-Klein scenario, spinorial fields show a natural chirality, as it can be easily understood after a first glance to Dirac equation in 5 dimensions. $\gamma^{5}$ can be used as a good Dirac matrix, for it obeys Dirac algebra. If we substitute the 4-dimensional identity $I=P_{R} + P_{L}$ in (\ref{13}),$P_{R}\equiv \frac{I+\gamma^{5}}{2}$ and $P_{L}\equiv \frac{I-\gamma^{5}}{2}$ being the right and left projection operators, respectively, we find different Dirac equation for the two chiral states when varying, for example, with respect to $\bar{\chi}_{i}$:
\begin{equation}\label{chiraldestra}
-\gamma^{\bar{\mu}}\partial_{\bar{\mu}}\psi_{iR}+i\frac{2\pi}{L}N_{iR}\psi_{iR}=0,
\end{equation}
\begin{equation}\label{chiralsinistra}
\gamma^{\bar{\mu}}\partial_{\bar{\mu}}\psi_{iL}+i\frac{2\pi}{L}N_{iL}\psi_{iL}=0;
\end{equation}
if we want the parameter $N_{i}$ to be related with the weak hyper-charge value, we cannot assume $N_{iR}=-N_{iL}$, for it would be a kind of particle/antiparticle condition, and we have to define ab initio two different fields:
\begin{equation}\label{pr}
{^{5}\chi}_{iR}(x^{\rho},x^{5})=\frac{1}{\sqrt{L}}{^{4}P_{R}\psi}_{i}(x^{\rho})e^{\frac{i2\pi N_{iR}x^{5}}{L}}=\frac{1}{\sqrt{L}}{^{4}\psi}_{iR}(x^{\rho})e^{\frac{i2\pi N_{iR}x^{5}}{L}}
\end{equation}
\begin{equation}
{^{5}\chi}_{iL}(x^{\rho},x^{5})=\frac{1}{\sqrt{L}}{^{4}P_{L}\psi}_{i}(x^{\rho})e^{\frac{i2\pi N_{iL}x^{5}}{L}}=\frac{1}{\sqrt{L}}{^{4}\psi}_{iL}(x^{\rho})e^{\frac{i2\pi N_{iL}x^{5}}{L}};
\end{equation}
it's interesting to notice that the two chirality states perfectly fit one of the most striking features of the Standard Electro-Weak Model, i.e. left/right asymmetry, also as far as non Abelian local SU(2) transformations are concerned.\\   
Neglecting, for the moment, gauge and spin connections, whose role will be thoroughly developed in the following sections, we can check the previous hypothesis about the weak hyper-charge geometrization.\\ 
For a generic matter field, the pertinent Lagrangian density,from (\ref{festa}),after dimensional reduction, reads
\begin{equation}\label{pezzo}
{^{4}L}=-\frac{i}{2}\left[... -\bar{\psi_{j}}\gamma^{\mu}\tilde{g}'k\tilde{B}_{\mu}(\frac{i2\pi}{L}N_{j})\psi_{j}\right] + H.C.+...;
\end{equation}\\
first, we can identify the weak hyper-charge coupling constant $g'$
\begin{equation}\label{g'}
g'\equiv \tilde{g}'\frac{2\pi k}{L}
\end{equation}
where the functional dependence on the length of the ring is common in Kaluza-Klein theories.\\
The real gauge field $B_{\mu}$ is found by affixing the new parameter $M$,
\begin{equation}\label{emme} 
B_{\mu}\equiv \frac{\tilde{B}_{\mu}}{M},
\end{equation}
so that the Standard Model is restored by the request 
\begin{equation}\label{y}
MN_{i}\equiv -y_{i},
\end{equation}
$y_{i}$ being the weak hyper-charge quantum number of the field $\chi_{i}$; since $N_{i}\in Z$, without loss of generalities we can choose $M=\frac{1}{6}$, so that the periodicity is an integer sub-multiple of the circumference $L$. Because of (\ref{emme}), for the gauge transformation (\ref{alfa}) we have  $\alpha(x^{\nu}) \rightarrow M\alpha(x^{\nu})$, and the bosonic action (\ref{azionefinalmentedue}) becomes
\begin{equation}
S=...-\frac{c^{3}k^{2}\tilde{g}'^{2}M^{2}}{64\pi G}\int d^{4}x\sqrt{-g}F^{\mu\nu}F_{\mu\nu},
\end{equation}
so that we can estimate the length of the extra-ring: with (\ref{g'}) and making use of the relation $e=g' \cos\theta_{W}$, which relates the electric charge and $g'$, we get $L\approx8.86\cdot10^{-30}$.\\
Conserved quantities can be defined for translations along the extra-ring \cite{fra}: the stress-energy tensor is extended to 5 dimensions as
\begin{equation}
T_{AB}=-\eta_{AB}L+\Sigma_{r}\frac{\partial L}{\partial(\partial^{A} \psi_{r})}\partial_{B}\psi_{r},
\end{equation}
and it obeys the conservation equation
\begin{equation}
\partial_{A} T^{AB}=0,
\end{equation}
so that the conserved momentum components are
\begin{equation}
cP_{A}=\int d^{3}x dx^{5} T_{0A};
\end{equation}
for the fifth component we have
\begin{align}
P_{5}= \int d^{3}x dx^{5} \left[-\frac{i\hbar}{2}\overline{\chi_{j}}\gamma^{0}\partial_{5}\chi_{j}+H.C. \right]=\nonumber\\
=\int d^{3}x dx^{5}\left[ \frac{i\hbar}{2}\chi_{j}^{+}\frac{i2\pi N_{j}}{L}\chi_{j}+H.C.\right]
=i\hbar \frac{i2\pi N_{j}}{L} \int d^{3}\psi_{j}^{+}\psi_{j},
\end{align}
and, since it is a conserved quantity, it can be multiplied times an arbitrary constant in order to get the right dimension of a charge: with (\ref{g'}) and (\ref{y}), we obtain therefore that the fifth momentum component is the weak hyper-charge $Q_{Y}$
\begin{equation}
P_{5}=Q_{Y}=g'y_{i}\int d^{3}\psi_{j}^{+}\psi_{j}.
\end{equation}
The identification of the constant of motion $P_{5}$ to the conserved weak hyper-charge $Q_{Y}$ is allowed by the choice of a complex phase dependence on the fifth coordinate of the wave function (\ref{chi}), which is coherent, from a quantum-mechanical point of view, with the representation of an eigen-state of $x^{5}$.\\
It is possible to verify that  spinorial fields (\ref{chi}) meet the correct weak hyper-charge transformation law, generated, in this model, by the second of (\ref{ammesse}):
\begin{equation}\label{5primo}
\chi_{i}\rightarrow \chi_{i}'=\frac{1}{\sqrt{L}}e^{\frac{i2\pi N_{i}}{L}(x'^{5}-Mk\tilde{g}'\alpha'(x^{\rho})))}{^{4}\psi}_{i}(x^{\rho})\equiv
\frac{1}{\sqrt{L}}e^{\frac{i2\pi}{L}N_{i}x'^{5}}{^{4}\psi}';
\end{equation}
in fact, (\ref{emme}) implies $\alpha(x^{\nu}) \rightarrow M\alpha(x^{\nu})$, and the right gauge transformation 
\begin{equation}
\psi \rightarrow \psi' =e^{ig'y_{i}\alpha'(x^{\nu})}\psi
\end{equation}
is achieved, where we have used (\ref{g'}) and (\ref{y}). It has just been shown, therefore, that U(1) Abelian gauge transformations can be ascribed, in a 5-dimensional scenario, to the cylindricity hypothesis, from which (\ref{ammesse}) are derived, and the factorized expression (\ref{chi}) is found to be a good extension: (\ref{chi}) does not break the cilindricity hypothesis itself, as spinorial fields are not physical objects. \\
Furthermore, it's worth remarking that the apparently different phase factor defined in (\ref{5primo}) is unimportant, for the right and the left side have the same periodicity $L$, and it does not modify expectation values of physical observables, that are quadratic functions of the fields.

\section{4-dimensional local Lorentz group}
In General Relativity, a generic infinitesimal coordinate transformation reads
\begin{equation}\label{diff}
x^{\mu}\rightarrow x'^{\mu}(x)=x^{\mu}+\xi^{\mu}(x),
\end{equation}
while an infinitesimal local Lorentz transformation is given by
\begin{equation}
x^{\mu}\rightarrow x'^{\mu}(x)=\Lambda^{\mu}_{\ \nu}x^{\nu}=x^{\mu}+\epsilon^{\mu}_{\ \nu}(x)x^{\nu};
\end{equation}
the latter can be easily expressed in terms of the former by
\begin{equation}\label{ops}
x^{\mu}\rightarrow x'^{\mu}(x)=x^{\mu}+\epsilon^{\mu}_{\ \nu}(x)x^{\nu}=x^{\mu}+\tilde{\epsilon}^{\mu}(x).
\end{equation}
If we introduce interaction with matter, this is not true, as the transformation law of fermionic fields under local Lorentz transformations cannot be compared to that of a generic tensor under general coordinate transformation: fermionic fields are described by a spinorial representation of the Lorentz group, while the diffeomorphism group has none. This issue suggests to treat local Lorentz transformations like an independent gauge field by defining a covariant derivative and a bosonic Lagrangian density, \cite{mermon}. It can be generalized to a flat 4-dimensional manifold, endowed with a set of tetradic vectors $e^{\bar{a}}_{\mu}$: the Lagrangian density reads
\begin{equation}
L=-\frac{i}{2}\bar{\psi}\gamma^{\bar{a}}e^{\mu}_{\bar{a}}\partial_{\mu}\psi +H.C.,
\end{equation}
where invariance under diffeomorphism is assured by the tetrad.Invariance under local Lorentz transformations
\begin{equation}\label{in}
\psi(x)\rightarrow \psi'(x')=S(\Lambda)(x)\psi(x),\bar{\psi}(x)\rightarrow \bar{\psi}'(x')=\bar{\psi}(x)S^{-1}(\Lambda)(x),
\end{equation}
\begin{equation}
S(\Lambda)\gamma^{\bar{a}}S^{-1}(\Lambda)=\Lambda^{\bar{a}}_{\bar{b}}\gamma^{\bar{b}},
\end{equation}
where
\begin{equation}\label{sigma1}
S(\Lambda)(x)=I+\frac{1}{2}\epsilon^{\bar{a}\bar{b}}\Sigma_{\bar{a}\bar{b}},\qquad \Sigma^{\bar{a}\bar{b}}=\frac{i}{4}[\gamma^{\bar{a}},\gamma^{\bar{b}}],
\end{equation}
is restored by the covariant derivative
\begin{equation}\label{rondine}
D_{\bar{a}}\psi=e_{\bar{a}}^{\mu}D_{\mu}\psi=e_{\bar{a}}^{\mu}(\partial_{\mu}-C_{\mu})\psi,
\end{equation}
where $C_{\mu}=C^{\bar{a}\bar{b}}_{\mu}\Omega_{\bar{a}\bar{b}}$ transforms like
\begin{equation}\label{gauge}
C^{\bar{a}\bar{b}}_{\mu}\rightarrow S(\Lambda)(x)C^{\bar{a}\bar{b}}S^{-1}(\Lambda)(x)+S(\Lambda)(x)\partial_{\mu}S^{-1}(\Lambda)(x),
\end{equation}
and, for infinitesimal $\epsilon$,
\begin{equation}\label{quattro}
C^{\bar{a}\bar{b}}_{\mu}\rightarrow C^{\bar{a}\bar{b}}_{\mu}+\partial_{\mu}\epsilon^{\bar{a}\bar{b}}-D^{\bar{a}\bar{b}}_{\bar{c}\bar{d}\bar{e}\bar{f}}C^{\bar{c}\bar{d}}_{\mu}\epsilon^{\bar{e}\bar{f}},
\end{equation}
$\Omega_{\bar{a}\bar{b}}$ and$D^{\bar{a}\bar{b}}_{\bar{c}\bar{d}\bar{e}\bar{f}}$ being suitable generators and structure constants of the group, respectively.\\
In a curved space time, the validity of Dirac equation is ensured by the request that Dirac algebra be still valid even in the new non Minkoskian metric, i.e. the tetradic projection of Dirac matrices still obey
\begin{equation}
[\gamma^{\bar{a}},\gamma^{\bar{b}}]_{+}=2I\eta^{\bar{a}\bar{b}}.
\end{equation}
If we consider Riemann-Cartan spaces, endowed with the affine connection
\begin{equation}\label{cane}
\Gamma^{\mu}_{\nu\rho}=\left\{
\begin{array}{c}
\mu\\
\nu\rho
\end{array}
\right\}-K_{\nu\rho}^{\mu}
\end{equation}we look for an operator $D_{\mu}$ which allows
\begin{equation}
D_{\mu}\gamma_{\nu}=0;
\end{equation}
such an operator is found to be
\begin{equation}\label{gatto}
D_{\mu}A=\nabla_{\mu}A-[\Gamma_{\mu},A]
\end{equation}
for a generic geometrical object, and
\begin{equation}\label{canarino}
D_{\mu}\psi=\partial_{\mu}\psi-\Gamma_{\mu}\psi,\quad D_{\mu}\bar{\psi}=\partial_{\mu}\bar{\psi}+\bar{\psi}\Gamma^{\mu}
\end{equation}
for fermionic fields, so that the matter Lagrangian density reads
\begin{equation}
L_{M}=-\frac{i}{2}\bar{\psi}\gamma^{\bar{a}}e^{\mu}_{\bar{a}}D_{\mu}\psi +H.C..
\end{equation} 
By substitution of (\ref{cane}) in (\ref{gatto}), after standard manipulation one finds
\begin{equation}
\Gamma_{\mu}=\Gamma_{\mu}^{R}+\Gamma_{\mu}^{K}=
\frac{1}{2}R^{\bar{a}\bar{b}}_{\mu}\Sigma_{\bar{a}\bar{b}}+\frac{1}{2}A^{\bar{a}\bar{b}}_{\mu}\Sigma_{\bar{a}\bar{b}},
\end{equation}
where the tetradic projection of the Ricci coefficients and of the contortion field are defined, respectively,
\begin{equation}
R_{\bar{a}\bar{b}\mu}=R_{\bar{a}\bar{b}\bar{c}}e^{\bar{c}}_{\mu},
\end{equation}
\begin{equation}\label{cappa}
A_{\bar{a}\bar{b}\mu}\equiv -K_{\rho\sigma\mu}e^{\rho}_{\bar{a}}e^{\sigma}_{\bar{b}}.
\end{equation}\\
If we now consider flat spaces, i.e. $R_{\mu\nu\rho\sigma}\equiv 0$ and compare (\ref{rondine}), (\ref{canarino}) and(\ref{cappa}), we can identify
\begin{equation}\label{com}
C_{\bar{a}\bar{b}\mu} \equiv A_{\bar{a}\bar{b}\mu},\qquad \Omega_{\bar{a}\bar{b}}\equiv \frac{1}{2}\Sigma_{\bar{a}\bar{b}}:
\end{equation}we can find, therefore, gauge fields and generators of the local Lorentz group. Structure constants in (\ref{quattro}) can be evaluated from structure constants of the Lorentz group $F^{\bar{e}\bar{f}}_{\bar{a}\bar{b}\bar{c}\bar{d}}$:
\begin{equation}
[\Sigma_{\bar{a}\bar{b}},\Sigma_{\bar{c}\bar{d}}]=\eta_{\bar{c}[\bar{a}}\Sigma_{\bar{b}]\bar{d}}-\eta_{\bar{d}[\bar{a}}\Sigma_{\bar{b}]\bar{c}}=\left[ \eta_{\bar{c}[\bar{a}}\delta_{\bar{b}]}^{[\bar{e}}delta_{\bar{d}}^{\bar{f}]}-\eta_{\bar{d}[\bar{a}}\delta_{\bar{b}]}^{[\bar{e}}\delta_{\bar{c}}^{\bar{f}]}\right]\Sigma_{\bar{e}\bar{f}}\equiv
F^{\bar{e}\bar{f}}_{\bar{a}\bar{b}\bar{c}\bar{d}}\Sigma_{\bar{e}\bar{f}}
\end{equation}
so that $D^{\bar{e}\bar{f}}_{\bar{a}\bar{b}\bar{c}\bar{d}}=\frac{1}{2}F^{\bar{e}\bar{f}}_{\bar{a}\bar{b}\bar{c}\bar{d}}$.\\
We have now all the terms to explain how to identify a local Lorentz transformation to a gauge transformation: the main difficulty in this task is the fact that a gauge transformation is defined in the same point of the space time, i.e. $\psi(x)\rightarrow \psi'(x)$, while a Lorentz transformations isn't, i.e. $\psi(x)\rightarrow \psi'(x')$. The set of tetradic vectors plays an essential role in in solving this discrepancy, and the implications of (\ref{ops}) need further developments: because of local Lorentz transformation, a tetradic vector transforms like
\begin{equation}
e'^{\bar{a}}_{\mu}(x')= \Lambda^{\bar{a}}_{\ \bar{b}}(x')e^{\bar{a}}_{\mu}(x'),  
\end{equation}
whereas, for a generic diffeomorphism, (\ref{diff}),
\begin{equation}\label{new}
e^{\bar{a}}_{\mu}(x)\rightarrow e'^{\bar{a}}_{\mu}(x')=e^{\bar{a}}_{\mu}(x)\frac{\partial x^{\rho}}{\partial x'^{\mu}}\approx e^{\bar{a}}_{\mu}(x)+e^{\bar{a}}_{\rho}(x)\frac{\partial \xi^{\rho}}{\partial x'^{\mu}};
\end{equation}
the role of $\epsilon^{\bar{a}\bar{b}}$ in formulas (\ref{in})$\rightarrow$(\ref{quattro}) becomes therefore crucial in pulling the transformation back to the same point, as long as gauge transformations are concerned.\\
After straightforward calculation, one finds from (\ref{new})
\begin{equation}
e'^{\bar{a}}_{\mu}(x')=e^{\bar{a}}_{\mu}(x')+e^{\bar{b}}_{\mu}(x')\left[-\partial _{\bar{b}}\xi^{\bar{a}}+\lambda^{\bar{a}}_{\ \bar{c}\bar{b}}\xi^{\bar{c}}\right]\equiv
e^{\bar{a}}_{\mu}(x')+e^{\bar{b}}_{\mu}(x')\epsilon^{\bar{a}}_{\bar{b}},
\end{equation}
where
\begin{equation}\label{fred}
\epsilon^{\bar{a}}_{\ \bar{b}}\equiv -\partial _{\bar{b}}\xi^{\bar{a}}+\lambda^{\bar{a}}_{\ \bar{c}\bar{b}}\xi^{\bar{c}} \equiv -\partial _{\bar{b}}\xi^{\bar{a}}+R^{\bar{a}}_{\ \bar{c}\bar{b}}\xi^{\bar{c}}-R^{\bar{a}}_{\ \bar{b}\bar{c}}\xi^{\bar{c}}\equiv -D_{\bar{b}}\xi^{\bar{a}}-R^{\bar{a}}_{\ \bar{b}\bar{c}}\xi^{\bar{c}};
\end{equation}
here $\lambda_{\bar{a}\bar{b}\bar{c}}=R_{\bar{a}\bar{b}\bar{c}}-R_{\bar{a}\bar{c}\bar{b}}$ are the anolonomy coefficients.\\
The number of degrees of freedom for generic diffeomorphisms, 16,  and that for Lorentz transformations, 6, lead to the request that (\ref{diff}) be an isometry, i.e. $\nabla_{(\mu}\xi_{\nu)}=0$, so that in (\ref{fred}) only the anti-symmetric part of $D_{\bar{b}}\xi^{\bar{a}}$ doesn't vanish. Finally we get
\begin{equation}\label{freddy}
\epsilon_{\bar{a}\bar{b}}=D_{[\bar{a}}\xi_{\bar{b}]}-R_{\bar{a}\bar{b}\bar{c}}\xi^{\bar{c}},
\end{equation}
which is anti-symmetric, i.e. $\epsilon_{\bar{a}\bar{b}}=-\epsilon_{\bar{b}\bar{a}}$. Formula (\ref{freddy}) shows once more that $R_{\bar{a}\bar{b}\bar{c}}$ cannot be a gauge field, for it defines the transformation. \\   
Field equations can be calculated from variational principles. In a gauge theory, gauge connections are primitive objects,so that the total action $S \equiv S(e,A,\psi)$ reads
\begin{equation}\label{svaria}
S \equiv {S}(e,A,\psi) \equiv -\frac{1}{2c\chi}\int d^{4}x\sqrt{-g}[R(e)-2\chi (L_{M}-\frac{1}{4}E^{2}F_{\mu\nu}(A)F^{\mu\nu}(A))],
\end{equation}\\
where $E$ is an opportune coupling constant, whose value must be small compared to the energy scales of modern-day experiments, since the effects of the 4-dimensional local Lorentz group have not been revealed yet.\\
Variation with respect to tetradic vectors, $\delta e^{\bar{a}}_{\mu}$, leads to the tetradic projection of Einstein equations, with Yang-Mill tensor $T^{\mu\nu}$  as source
\begin{equation}
(R_{\mu\nu}-\frac{1}{2}g_{\mu\nu}R)e^{\nu}_{\bar{a}}=\chi T_{\mu\bar{a}}.
\end{equation}
Variation with respect to the gauge field $\delta A_{\mu}^{\bar{a}\bar{b}}$ brings Yang-Mill equation, the spinorial current density being the source:
\begin{equation}\label{bo}
D_{\mu}F^{\mu\nu}_{\bar{a}\bar{b}}=-J^{\nu}_{\bar{a}\bar{b}}.
\end{equation}
Variation with respect to the adjoint spinorial field $\delta \bar{\psi}$ gets Dirac equation for $\psi$
\begin{equation}
\gamma^{\mu}D_{\mu}\psi=0,
\end{equation}
and vice-versa.\\
Conserved quantities can be found by the comparison of local Lorentz transformations and gauge transformations. In particular, because the current density (\ref{bo}) obeys the conservation law
\begin{equation}
\partial_{\mu}J^{\mu \bar{a}\bar{b}}=0,
\end{equation}
we have the conserved (gauge) charge
\begin{equation}\label{mah}
Q^{0 \bar{a}\bar{b}}=\int d^{3}x J^{0 \bar{a}\bar{b}}=const;
\end{equation}
on the other hand, the conserved quantity for Lorentz transformations in flat space time is the angular momentum tensor $M^{\mu\nu}$: the tetradic projection of the spin term reads
\begin{equation}\label{mahh}
M^{\bar{a}\bar{b}}= \int  d^{3}x  \pi_{r}\Sigma^{\bar{a}\bar{b}}_{rs}\psi_{s}= const.,
\end{equation}
which coincides to (\ref{mah}), provided that $\pi_{r}$ is the  density of momentum conjugate to the field $\psi_{r}$, i.e. $\pi_{r}=\partial L / \partial \dot{\psi}_{r}$. The parameter $\epsilon^{\bar{a}\bar{b}}$ defined in (\ref{freddy}) renders this identification possible.

\section{5-dimensional Lorentz group and local SU(2)}
The results of the previous section can be generalized to a higher number of dimensions, if interactions other than gravity are to be geometrized. The proceeding, however, is not so linear, because of the structure of the manifold (\ref{struttura}) we are dealing with: a generic 5-dimensional Lorentz transformation reads
\begin{equation}
x^{\Omega}(x^{\mu},x^{5})\rightarrow x'^{\Omega}(x^{\mu},x^{5})=x^{\Omega}+\epsilon^{\Omega}_{\ \Pi}x^{\Pi},
\end{equation} 
but, if we want to define the Lorentz group for (\ref{struttura}), the set of transformations (\ref{ammesse}) leads to the constraint  $\epsilon^{\Omega 5}\equiv 0$.\\
The global 5-dimensional Lorentz group can be therefore described by the generators $\Sigma^{\Omega\Pi}=\frac{i}{4}[\gamma^{\Omega},\gamma^{\Pi}]$, where $\gamma^{5}$ is the fifth Dirac matrix, as shown in (15). The 5-dimensional Lagrangian density reads
\begin{equation}
L=-\frac{1}{2}\left[\bar{\chi}_{L}\gamma^{\bar{\Omega}}\partial_{\Omega}\chi_{L} + \bar{\chi}_{R}\gamma^{\bar{\Omega}}\partial_{\Omega}\chi_{R} \right] + H.C.,
\end{equation}
where the extra-dimensional derivatives vanish because of the properties of the right/\\left projection operators. Invariance under 5-dimensional global Lorentz transformations is assured by the transformation rules for the spinorial fields, $\chi\rightarrow S(\Lambda)\chi$, $\bar{\chi}\rightarrow \bar{\chi}S^{-1}(\Lambda)$, and for Dirac matrices, $S(\Lambda)\gamma^{\Omega}S^{-1}(\Lambda)=\Lambda^{\Omega}_{\ \Pi}\gamma^{\Pi}$: in particular, because of the constraint $\epsilon^{\Omega 5}\equiv 0$, the 5-dimensional transformations lead to the ordinary 4-dimensional transformation, where the generator $\Sigma^{\mu 5}$ doesn't play any role.\\
When 5-dimensional local Lorentz transformations are considered, the previous analysis changes drastically, as shown in the second section: 5-dimensional local Lorentz transformations cannot be distinguished from generic 5-dimensional diffeomorphisms in the physical space time, while, in the tangent bundle, spinorial fields are the only geometrical object that can still experience the difference.\\
In the tangent bundle, as shown previously, the parameter $\epsilon^{\bar{A}\bar{B}}$ defines the transformation rules for spinorial fields and for Dirac matrices\footnote{We stress that the transformation law for the 4-dimensional Dirac matrices is violated by a term proportional to $gE\epsilon_{\bar{i}\bar{5}}\gamma^{j}\sigma_{i}$, as far as spinorial indices are concerned, as it will become clearer from (77) and (78); the fact that the effects of this term have not been observed yet can be explained by the small value of the constant $E$. The violation due to the transformation of world indices, proportional to $\epsilon^{\bar{a}}_{\ \bar{5}}\gamma^{\bar{5}}$ vanishes because of the properties of $\gamma^{\bar{5}}$. }, and, from (60), it can be shown that, for pure extra-dimensional coordinate transformations,
\begin{equation}\label{strano}
\epsilon_{\bar{a}\bar{5}}=\frac{1}{2}e^{\mu}_{\bar{a}}\partial_{\mu}\alpha(x^{\rho}):
\end{equation}
it must not be surprising that extra-dimensional transformations define both weak hyper-charge and weak isospin gauge transformations. Besides, equation (\ref{strano}) describes a different breaking as far as the physical space and the tangent bundle are concerned: the physical space is defined as (\ref{struttura}), while the tangent bundle is depicted by $\epsilon^{\bar{a}\bar{5}}$. Since $\epsilon^{\bar{a}\bar{5}}$ doesn't vanish, generators $\Sigma^{\bar{a}\bar{5}}$ recoup their own role; because invariance under Lorentz transformations is not determined by the explicit form of such generators, i.e. they act like further degrees of freedom, however, it is possible to choose their expression in order break the 5-dimensional group in the 4-dimensional group and a SU(2) group, as it can be inferred by studying the structure of 4-dimensional space rotations, \cite{jo}. It can be achieved by imposing that these generators commute with those of the ordinary 4-dimensional local group, \cite{mand}, as it will be shown throughout this section.\\
The 5-dimensional representation of the local Lorentz group cannot be, therefore, the immediate extension of the second of (\ref{sigma1}), that would be
\begin{equation}\label{5nonrotto}
\Sigma^{\bar{A}\bar{B}}=\frac{i}{4}[\gamma^{\bar{A}},\gamma^{\bar{B}}];
\end{equation}
while for $\bar{a}=0,1,2,3,$ the four- dimensional group is regained, $\Sigma^{\bar{i}\bar{5}}$ does not provide any known group, nor local $SU(2)$ non Abelian group, which we wish to describe. In fact, $\bar{A}=\bar{5}$ cannot be treated like $\bar{a}=0,1,2,3,$ because the fifth dimensional ring $S^{1}$ is just directly summed to $V^{4}$. Furthermore, $\Sigma^{\bar{0}\bar{5}}$ would be a redundant degree of freedom, for Abelian group U(1) has already been geometrized by a Kaluza-Klein scheme: $\Sigma^{\bar{0}\bar{5}}$ must be a vanishing quantity. The corresponding gauge field must vanish as well.\\
On the other hand, one has to consider that $\Sigma^{\bar{i}\bar{5}}$ is strictly connected to the extra-dimension, which is responsible for the chirality matter fields show in 5 dimensions, (\ref{chiraldestra}) (\ref{chiralsinistra}): it is, therefore, coherent to hypothesize an asymmetric right/left behavior for such generators, just like (\ref{5nonrotto}) would have in a generic manifold $V^{5}$, because of the properties of $\gamma^{5}$, $P_{R}$ and $P_{L}$.\\ 
In order to find the right expression for $\Sigma^{\bar{i}\bar{5}}$, we can start by extending (\ref{svaria}) to five dimensions, and then imposing that the components of the conserved current be the isospin conserved current. SU(2) local isospin group operates on left-handed isospin doublets, while, so far, we only have introduced ''singlets''; the right/left asymmetry we have just requested can be put into effect, first of all, by assuming that only left-handed fields undergo this transformations, i.e. right-handed fields are in the null-space of the corresponding generator: because of (\ref{pr}) and the properties of the projectors, we have $\Sigma^{\bar{i}\bar{5}} \propto P_{L}$.\\
Besides, we have to face the difficulty of introducing doublets; weak isospin doublets can be introduced by
\begin{equation}\label{doppiettichilep}
X_{f_{i}L} \equiv
\left(
\begin{array}{c}
\chi_{\nu_{i}L}\\
\chi_{l_{i}L}
\end{array}
\right)\equiv \frac{1}{\sqrt{L}}\Psi_{iL}e^{\frac{i2\pi}{L}N_{iL}x^{5}}:
\end{equation}\\
a different choice would describe the same physics, but matter fields should be relabeled, so (\ref{doppiettichilep}), and the equivalent for quark doublets, is a valid option. Even though it could sound forced to introduce doublets in a theory where no doublets are needed, we can remark that also right-handed fields can be arranged this way, but there would be no point in it, for the matrix operating on such doublets would only be the identity: the geometric 4-dimensional covariant derivative and weak hyper-charge acting on spinors, such doublets would naturally split up.\\
Suitable generators $\Sigma^{\bar{i}\bar{5}}$ can be finally deduced from (\ref{svaria}),by identifying them to the operators in the current term coupled to the gauge field.\\
In the Electro-Weak Model, the weak isospin current and the bosonic Lagrangian density are found to be, respectively
\begin{equation}
J^{\mu}_{i}=\frac{i}{2}\bar{\Psi}_{L}\gamma^{\mu}\sigma_{i}\Psi_{L},
\end{equation}
\begin{equation}
L=-\frac{1}{4}F_{\mu\nu}^{i}(W)F^{\mu\nu}_{i}(W)
\end{equation}
$\sigma_{i}$ being a Pauli matrix, while, from (\ref{svaria}), we have
\begin{equation}\label{mnbv}
{^{5}J}^{\bar{A}\bar{B}\mu}=\frac{1}{2}\bar{X}_{Lr}\gamma^{\mu}\Sigma^{\bar{A}\bar{B}}_{rs}X_{Ls}=\frac{1}{2}\bar{X}_{Lr}\gamma^{\mu}\Sigma^{\bar{a}\bar{b}}_{rs}X_{Ls}+\bar{X}_{Lr}\gamma^{\mu}\Sigma^{\bar{i}\bar{5}}_{rs}X_{Ls}
\end{equation} 
\begin{equation}\label{aereo}
{^{5}L}=-\frac{1}{4}E^{2}F_{\Omega\Pi}^{\bar{A}\bar{B}}F^{\Omega\Pi}_{\bar{A}\bar{B}}=-\frac{1}{4}E^{2}F_{\mu\nu}^{\bar{a}\bar{b}}(A)F^{\mu\nu}_{\bar{a}\bar{b}}(A)-2\frac{1}{4}E^{2}F_{\mu\nu}^{\bar{i}\bar{5}}(A)F^{\mu\nu}_{\bar{i}\bar{5}}(A).
\end{equation}
It is therefore easy to determine the suitable generators :
\begin{equation}\label{bvc}
\Sigma^{\bar{a}\bar{b}}\equiv \frac{i}{4}[\gamma^{\bar{a}},\gamma^{\bar{b}}],
\end{equation}  
\begin{equation}\label{bvcx}
\Sigma^{\bar{i}\bar{5}}_{rs}\equiv -ig\sqrt{2}E\frac{\sigma^{i}_{rs}}{2}P_{L};
\end{equation}
\begin{equation}\label{nullo}
\Sigma^{\bar{0}\bar{5}}_{rs}\equiv 0,
\end{equation}
as discussed above.\\
Structure constants ${^{5}C}^{\bar{A}\bar{B}}_{\bar{C}\bar{D}\bar{E}\bar{F}}$ are defined as
\begin{equation}\label{asdf}
[{^{5}\Omega}_{\bar{C}\bar{D}},{^{5}\Omega}_{\bar{E}\bar{F}}]={^{5}C}^{\bar{A}\bar{B}}_{\bar{C}\bar{D}\bar{E}\bar{F}}{^{5}\Omega}_{\bar{A}\bar{B}},
\end{equation}
where $ D^{\bar{a}\bar{b}}_{\bar{c}\bar{d}\bar{e}\bar{f}}$ are given by (\ref{quattro}). Since $\Sigma^{\bar{a}\bar{b}}$ and $\Sigma^{\bar{i}\bar{5}}$ operate on different spaces, and $\Sigma^{\bar{0}\bar{5}}$ vanishes, the only non-vanishing commutators are, because of (\ref{com}), (\ref{bvc}), (\ref{bvcx}), (\ref{nullo}) and (\ref{sigma1}), 
\begin{equation}\label{strut1}
[\Sigma_{\bar{a}\bar{b}},\Sigma_{\bar{c}\bar{d}}]=D^{\bar{e}\bar{f}}_{\bar{a}\bar{b}\bar{c}\bar{d}}\Sigma_{\bar{a}\bar{b}}
\end{equation}
\begin{equation}\label{strut2}
[\Sigma_{\bar{i}\bar{5}},\Sigma_{\bar{j}\bar{5}}]=C^{\bar{k}\bar{5}}_{\bar{i}\bar{5}\bar{j}\bar{5}}\Sigma_{\bar{k}\bar{5}},
\end{equation}
so that 
\begin{equation}\label{ijk}
C^{\bar{k}\bar{5}}_{\bar{i}\bar{5}\bar{j}\bar{5}}\equiv g\sqrt{2}E\epsilon_{ijk}.
\end{equation}
From (\ref{asdf}),(\ref{strut1}) and (\ref{strut2}), we can see that the total group we are dealing with is the direct sum of the ordinary 4-dimensional Lorentz group and the weak isospin group: the structure
\begin{equation}\label{mandula}
SO(3,1)\otimes SU(2)
\end{equation} 
is expressed by the vanishing commutators
\begin{equation}
[\Sigma_{\bar{a}\bar{b}},\Sigma_{\bar{i}\bar{5}}]=0,
\end{equation}
i.e. the former $\Sigma^{\bar{A}\bar{B}}$ is broken in the Lorentz group and an internal symmetry group.\\
So far, we can check that the fields $A^{\bar{A}\bar{B}}_{\Omega}$ undergo the standard world and gauge transformations. For $A^{\bar{A}\bar{B}}_{\mu}$'s fields, the former reads
\begin{equation}
A^{\bar{A}\bar{B}}_{\mu} \rightarrow A'^{\bar{A}\bar{B}}_{\mu}= A^{\bar{A}\bar{B}}_{\Omega} \frac{\partial x^{\Omega}}{\partial x'^{\mu}}=A^{\bar{A}\bar{B}}_{\nu} \frac{\partial x^{\nu}}{\partial x'^{\mu}}+A^{\bar{A}\bar{B}}_{5} \frac{\partial x^{5}}{\partial x'^{\mu}}:
\end{equation}
for the second of (\ref{ammesse}), $\frac{\partial x^{5}}{\partial x'^{\mu}}$ does not vanish automatically, and we need to impose 
\begin{equation}
A^{\bar{A}\bar{B}}_{5}\equiv 0
\end{equation}
in order to regain the ordinary 4-dimensional word transformation
\begin{equation}
A^{\bar{A}\bar{B}}_{\mu} \rightarrow A^{\bar{A}\bar{B}}_{\nu} \frac{\partial x^{\nu}}{\partial x'^{\mu}},
\end{equation}
i.e. $A^{\bar{A}\bar{B}}_{\mu}$'s fields behave like 4-vectors.\\
Gauge transformations are achieved by the immediate 5-dimensional extension of (\ref{quattro}), and (\ref{freddy}) must be taken into account:
\begin{equation}
A^{\bar{A}\bar{B}}_{\mu}\rightarrow A^{\bar{A}\bar{B}}_{\mu}+\partial_{\mu}\bar{\epsilon}^{\bar{A}\bar{B}}-D^{\bar{A}\bar{B}}_{\bar{C}\bar{D}\bar{E}\bar{F}}A^{\bar{C}\bar{D}}_{\mu}\epsilon^{\bar{E}\bar{F}}:
\end{equation}
because of (\ref{strut1}) and (\ref{strut2}), 4-dimensional local Lorentz transformations , as they were defined in (\ref{quattro}), are straight away restored, while for $A^{\bar{i}\bar{5}}_{\mu}$ we have
\begin{equation}\label{poi}
A^{\bar{i}\bar{5}}_{\mu}\rightarrow A^{\bar{i}\bar{5}}_{\mu}+\partial_{\mu}\bar{\epsilon}^{\bar{i}\bar{5}}-C^{\bar{i}\bar{5}}_{\bar{j}\bar{5}\bar{k}\bar{5}}A^{\bar{j}\bar{5}}_{\mu}\epsilon^{\bar{k}\bar{5}},
\end{equation}
where (\ref{strano}) has been taken into account.\\
Structure constants ensure that different gauge fields don't mix up.\\
If we consider (84), disregarding boosts, the 5-dimensional space rotation group is SO(4), while, in the Kaluza-Klein scheme, we have 
\begin{equation}\label{ora}
SO(4)\rightarrow SO(3)\otimes SU(2);
\end{equation}
no sooner boosts are considered than (\ref{mandula}) is reestablished, as $\bar{\epsilon}^{\bar{0}\bar{5}}$ is not experienced by spinorial fields: because $\Sigma^{\bar{0}\bar{5}}\equiv 0$ and $A^{\bar{0}\bar{5}}_{\mu}\equiv 0$, the parameter $\bar{\epsilon}^{\bar{0}\bar{5}}$, even though a non vanishing quantity, is not involved in any transformation. \\ 
This section is aimed to investigate how SU(2) weak isospin symmetry can be restored, so we can 
guess from now on
\begin{equation}\label{w}
W^{i}_{\mu}\equiv \sqrt{2}EA^{\bar{i}\bar{5}}_{\mu}:
\end{equation}
since $\Sigma^{\bar{i}\bar{5}}$ is always coupled to $A^{\bar{i}\bar{5}}_{\mu}$, we can put
\begin{equation}\label{sigma}
\Sigma^{\bar{i}\bar{5}}\rightarrow -ig\frac{\sigma^{i}}{2}P_{L},
\end{equation}
\begin{equation}\label{ep}
C^{\bar{i}\bar{5}}_{\bar{j}\bar{5}\bar{k}\bar{5}}\rightarrow g\epsilon_{ijk},
\end{equation}
\begin{equation}\label{eps}
\epsilon^{\bar{i}\bar{5}}\rightarrow \sqrt{2}E \epsilon^{\bar{i}\bar{5}}.
\end{equation}
This way there is no misunderstanding in (\ref{w}), (\ref{sigma}), (\ref{ep}), (\ref{eps}), since $\bar{i}$ index refers to the tangent bundle only, and $\bar{5}$ can be neglected, as it behaves like a passive index (it is not involved in any transformation). Equation (\ref{poi}) becomes
\begin{equation}
W^{i}_{\mu}\rightarrow W'^{i}_{\mu}=W^{i}_{\mu}+\partial_{\mu}\epsilon^{i} -g\epsilon_{ijk}W^{j}_{\mu}\epsilon^{k},
\end{equation}
where we have identified the weak isospin gauge field,(\ref{w}).
The bosonic Lagrangian density reads, finally,
\begin{equation}\label{elicottero}
L=-\frac{1}{4}E^{2}F_{\mu\nu}^{\bar{a}\bar{b}}(A)F^{\mu\nu}_{\bar{a}\bar{b}}(A)-\frac{1}{4}F_{\mu\nu}^{i}(W)F^{\mu\nu}_{i}(W), 
\end{equation}
where
\begin{equation}
F^{i}_{\mu\nu}(W)=\partial_{\nu}W^{i}_{\mu}-\partial_{\mu}W^{i}_{\nu}+g\epsilon_{ijk}W^{j}_{\mu}W^{k}_{\nu}.
\end{equation}
4-dimensional Lorentz transformations and SU(2) gauge transformations are also restored for spinorial fields: the 5-dimensional transformation reads
\begin{equation}
\chi_{i}\rightarrow \chi'_{i}=e^{\Omega_{\bar{A}\bar{B}}\epsilon^{\bar{A}\bar{B}}},
\end{equation}
so that, for the first of (\ref{ammesse}), we have
\begin{align}
\chi_{i}= {^{5}\chi}_{i}(x^{\rho},x^{5})=\frac{1}{\sqrt{L}}{^{4}\psi}_{i}(x^{\rho})e^{\frac{i2\pi N_{i}x^{5}}{L}}\rightarrow\nonumber\\ \rightarrow\chi'_{i}=e^{\Omega_{\bar{a}\bar{b}}\epsilon^{\bar{a}\bar{b}}}{^{5}\chi}_{i}(x^{\rho},x^{5})=\frac{1}{\sqrt{L}}e^{\Omega_{\bar{a}\bar{b}}\epsilon^{\bar{a}\bar{b}}}{^{4}\psi}_{i}(x^{\rho})e^{\frac{i2\pi N_{i}x^{5}}{L}}:
\end{align} 
this way
\begin{equation} 
{^{4}\psi}_{i}\rightarrow {^{4}\psi}'_{i}=e^{\frac{1}{2}\Sigma^{\bar{a}\bar{b}}\epsilon_{\bar{a}\bar{b}}}{^{4}\psi}_{i}.
\end{equation}\\
For the second of (\ref{ammesse}), on the other hand, we get
\begin{align}
X_{iL}= {^{5}X}_{iL}(x^{\rho},x^{5})=\frac{1}{\sqrt{L}}{^{4}\Psi}_{iL}(x^{\rho})e^{\frac{i2\pi N_{iL}x^{5}}{L}}\rightarrow\nonumber\\ \rightarrow X'_{iL}=e^{2\Omega_{\bar{i}\bar{5}}\epsilon^{\bar{i}\bar{5}}}{^{5}X}_{iL}(x^{\rho},x'^{5})=\frac{1}{\sqrt{L}}e^{\frac{i}{2}g\sigma_{i}\epsilon^{i}}e^{\frac{i2\pi N_{iL}x'^{5}}{L}}{^{4}\Psi}_{iL}(x^{\rho}):
\end{align}
the 4-dimensional spinor transforms like
\begin{equation}\label{whole}
{^{4}\Psi}_{iL} \rightarrow {^{4}\Psi}'_{iL}=e^{ig'y_{i}\alpha'(x^{\nu})}e^{\frac{i}{2}g\sigma_{i}\epsilon^{i}}{^{4}\Psi}_{iL},
\end{equation}
where, according to (\ref{w}), (\ref{sigma}) and the results of the first section.
Weak hyper-charge and weak isospin transformations are here generated by the same coordinate transformation, i.e. it's impossible to generate the former without the latter , and vice-versa: in the Standard Model, however, this is the complete spinorial field transformation, and, as far as the bosonic fields $A_{\mu}$, $Z_{\mu}$, $W_{\mu}$ and $W^{+}_{\mu}$ are concerned, it is not possible to distinguish between the different transformations individually.\\
Conserved quantities can be found in 5 dimensions as well: according to the considerations which led to (\ref{mah}) and (\ref{mahh}), it's possible to establish conserved charges because of the breaking of the 5-dimensional Lorentz group. Since the current density (\ref{mnbv}) still obeys a conservation law
\begin{equation}
\partial_{\Omega}J^{\Omega}_{\bar{A}\bar{B}}=0,
\end{equation}
we get the 5-dimensional conserved charge
\begin{equation}\label{mah2}
Q^{0 \bar{A}\bar{B}}=\int^{L}_{0} d^{3}xdx^{5}J^{0 \bar{A}\bar{B}}=const.
\end{equation}
The spin angular momentum tensor can be extended to 5 dimensions, too, and its tetradic projection reads
\begin{equation}
M^{\bar{A}\bar{B}}= \int ^{L}_{0} d^{3}x dx^{5} {^{5}\pi}_{r}\Sigma^{\bar{A}\bar{B}}_{rs}\chi_{s}= const.,
\end{equation}
which coincides again to (\ref{mah2}), and ${^{5}\pi}_{r}$ is the 5-dimensional density of momentum conjugate to the field $\chi_{r}$.\\
For weak isospin transformations, weak isospin charge is just the tetradic projection of the extra-dimensional part of (\ref{mah2}):
\begin{equation}
Q^{0}_{i}=M^{\bar{i}\bar{5}}=\int ^{L}_{0} d^{3}x dx^{5} {^{5}\pi}_{r}\Sigma^{\overline{i}\overline{5}}_{rs}X_{s}=\frac{-ig}{2}\int d^{3}x\Psi^{+}_{L}\sigma_{i}\Psi_{L},
\end{equation} where we have used the results worked out in this section. Once again, the role of the indices $\bar{i}$ and $\bar{5}$ allows us to better understand the breaking of the 5-dimensional Lorentz group into a 4-dimensional Lorentz group and an internal symmetry.

\section{Restoration of the Electro-Weak Model}

We can develop now the whole Lagrangian density (\ref{festa}) and substitute the covariant derivative (\ref{canarino}) in 5 dimensions.\\
The tetradic projection of the connection is
\begin{equation}
\Gamma_{\bar{A}}=e_{\bar{A}}^{\Omega}\Gamma_{\Omega}:
\end{equation}
for the gauge connections \cite{gio}, \cite{mar}, we have
\begin{equation}
\Gamma^{K}_{\bar{\mu}}=\frac{1}{2}\Sigma_{\bar{A}\bar{B}}A_{\mu}^{\bar{A}\bar{B}}=\frac{1}{2}\Sigma_{\bar{a}\bar{b}}A_{\mu}^{\bar{a}\bar{b}}+\frac{1}{2}\Sigma_{\bar{i}\bar{5}}A_{\mu}^{\bar{i}\bar{5}}
\end{equation}
and
\begin{equation}
\Gamma^{K}_{\bar{5}}\equiv 0,
\end{equation}
while for spin connections some more investigation is in order. In fact, for the manifold (\ref{struttura}), we get no connection for translations, as far as $S^{1}$ is concerned, and 4-dimensional connections are found to be purely Riemannian, so that
\begin{equation}
\Gamma^{R}_{\bar{\mu}}\equiv {^{4}\Gamma}_{\bar{\mu}},
\end{equation}
and
\begin{equation}
\Gamma^{R}_{\bar{5}}\equiv 0.
\end{equation}\\
These spin connections follow naturally from the definition of (\ref{struttura}); if, instead of this structure ,we considered the group (\ref{5nonrotto}), \cite{bla}, non-minimal coupling terms would arise from the tetradic projection of spin connection.\\ 
Since matter fields are introduced in their chiral states, extra-dimensional derivatives vanish when coupled to $\gamma^{\bar{5}}$, according to the projection operators' properties and to (\ref{gamma}):
\begin{equation}\label{tom}
\bar{\chi}_{a}\gamma^{\bar{5}}\partial_{\bar{5}}\chi_{a}\equiv 0
\end{equation}
where $a$ refers to right-handed singlets and to left-handed doublets. The property (\ref{tom}) corroborates the choice to introduce mass-less spinorial fields in their chirality states. In fact, if spinorial fields were introduced in a non-chiral form and then ''projected'' along the two different chiralities, besides the difficulty in defining such projection operators because of the values of the parameter $N_{i}$, mass-like terms would occur: these terms would break the symmetries of the model, just as in the Standard electro-weak theory, and, moreover, different masses would be implied for right- and left-handed fields in five dimensions; furthermore, these mass-like terms would vanish after dimensional reduction. These inconsistencies are the reasons why spinorial fields are to be established in the two different chirality states, and why they need to be mass-less, mass-like terms being generated by some spontaneous symmetry breaking mechanism.\\
Collecting all the terms together, we get
\begin{equation}
{^{5}L}=-\frac{i}{2} \left[\sum_{i}\bar{X}_{iR}\gamma^{\bar{A}}D_{\bar{A}}X_{iR} +\sum_{j}\bar{\chi}_{jR}\gamma^{\bar{A}}D_{\bar{A}}\chi_{jR}\right] + H.C.,
\end{equation}
where summation index $i$ runs over the three families of leptons and the three families of quarks, while $j$ runs over the six leptons and the six quarks. More explicitly,
\begin{align}
\nonumber
^{5}L= & -\frac{i}{2}\left[\sum_{i}\bar{X}_{iL}\gamma^{\bar{\mu}}(^{4}D_{\bar{\mu}}-\tilde{g}'k\tilde{B}_{\bar{\mu}}\frac{2i\pi N_{iL}}{L} +ig\frac{\hat{\sigma_{i}}}{2}W^{i}_{\bar{\mu}})X_{iL}+\right.
\\
& \left.+\sum_{j}
\bar{\chi}_{jR}\gamma^{\bar{\mu}}(^{4}D_{\bar{\mu}}-\tilde{g}'k\tilde{B}_{\bar{\mu}}\frac{2i\pi N_{jR}}{L}) \chi_{jR}\right]+H.C.
\end{align}
where
\begin{equation}
{^{4}D}_{\bar{\mu}}={^{4}\partial}_{\bar{\mu}}-{^{4}\Gamma}^{R}_{\bar{\mu}}-{^{4}\Gamma}^{K}_{\bar{\mu}};
\end{equation}
since there's no more ambiguity left, making use of (\ref{g'}), (\ref{emme}),  (\ref{y}), (\ref{bvcx}) and (\ref{w}), dimensional reduction can be carried out:
\begin{align}
\nonumber
{^{4}L}= &  -\frac{i}{2}\left[\sum_{i}\bar{\Psi}_{iL}\gamma^{\mu}(^{4}D_{\mu}-ig'B_{\mu}y_{iL}+ig\frac{\hat{\sigma}_{i}}{2}W^{i}_{\mu})\Psi_{iL}\right.
\\
& \left.+\sum_{j}\bar{\psi}_{jR}\gamma^{\mu}(^{4}D_{\mu}-ig'B_{\mu} y_{jR})\psi_{jR}\right]+H.C..
\end{align}
We can appreciate that the Electro-Weak Model has been restored thanks to the main assumptions of this work, i.e. the identification of the weak isospin gauge boson to the tetradic projection of the extra-dimensional contortion field, and the breaking of the previous 5-dimensional Lorentz group into the ordinary 4-dimensional Lorentz group and an internal symmetry. Dimensional reduction, which is intended to eliminate every object related to higher dimensions, is accomplished by the integration on the extra-coordinate: normalization factors and phase dependences on the fifth coordinate are therefore removed.

\section{Brief concluding remarks}

The geometrization of the Electro-Weak Model we've just accomplished has been carried out making use of a 5-dimensional Kaluza-Klein theory and of a gauge theory of the Lorentz group.\\
In the framework of a Kaluza-Klein theory it has been possible to geometrize U(1) weak hyper-charge symmetry: the gauge field has been introduced in the 5-dimensional metric tensor, and 4-dimensional chiral states have been defined by the splitting of the 5-dimensional Dirac equation, because of the 5-dimensional Dirac algebra.\\
The gauge theory of the local Lorentz group has been extended to 5 dimensions as far as the manifold (\ref{struttura}) involved in Kaluza-Klein schemes is concerned: extra-dimensional generators have been defined by the comparison of the extra-dimensional Lorentz conserved current to the weak isospin conserved current. The tetradic projection of the 5-dimensional spin part  of the angular momentum tensor has been found to be a conserved quantity. The character of spin connections has been explored through their tetradic projection, which were purely Riemannian.\\
In the Standard Model, unification is achieved by two different symmetry groups and two different coupling constants: this feature is here explained by the different geometrical origin of the two symmetry groups.  Two different properties of the 5-dimensional space-time, i.e. the metric tensor and the extra-dimensional contortion field, give rise to the transformations which spinorial an bosonic fields undergo: the different geometrical origin can explicate the reason why these groups are so formally dissimilar , and why coupling constants don't coincide. The chief peculiarity of electro-weak interactions, i.e. right/left asymmetry, is reinstated : it comes out naturally from the 5-dimensional Kaluza-Klein scheme, and is brought back in the whole transformation (\ref{whole}).\\
U(1) weak hyper-charge and SU(2) weak isospin transformations were generated from the same (extra-)coordinate transformation (\ref{ammesse}), so that the two gauge symmetries seem too strictly connected. In the Standard Electro-Weak Model, however, $B_{\mu}$ and $W^{3}_{\mu}$ are mixed up to form the observed gauge bosons $A_{\mu}$ and $Z_{\mu}$, so, as far as observed quantities are concerned, it is not possible to distinguish between the two transformations. Thus, the correlation which takes place in this theory is not so distant, in principle, from the Standard Model.\\
The description of the Standard Model we have developed fits mass-less fields only: mass-like terms would break the symmetries of the 5-dimensional model and would vanish after dimensional reduction. Higgs mechanism would bring the expected results, but it is not so clear how to define gauge connections for a scalar boson.

\section*{Acknowledgments}
We would like to thank Francesco Cianfrani for his helpful suggestions and comments on the subject.

\end{document}